\documentclass[aps,prb,twocolumn,amsmath,amssymb,showpacs]{revtex4-1}
\usepackage{amsfonts, hyperref, bm, times}
\usepackage{graphicx}
\pdfoutput=1

\begin{document}

\title{Magnetoelectric effect in topological insulator films beyond the linear response regime}

\author{Dashdeleg Baasanjav}
\affiliation{
  Institute for Materials Research,
  Tohoku University, Sendai 980-8577, Japan
}
\author{O.~A.~Tretiakov}
\affiliation{
  Institute for Materials Research,
  Tohoku University, Sendai 980-8577, Japan
}
\author{Kentaro Nomura}
\affiliation{
  Institute for Materials Research,
  Tohoku University, Sendai 980-8577, Japan
}

\date{June 16, 2013}

\begin{abstract}
We study the response of topological insulator films to strong magnetic and electric fields beyond the linear response theory. As a model, we use the three-dimensional lattice Wilson-Dirac Hamiltonian where we simultaneously introduce both magnetic field through the Peierls substitution and electric field as a potential energy depending on lattice coordinate. We compute the electron energy spectrum by numerically diagonalizing this Hamiltonian and obtain quantized magnetoelectric polarizability. In addition, we find that the magnetoelectric effect vanishes as the film width decreases due to the hybridization of surface wave functions. Furthermore, applying a gate voltage between the surfaces, we observe several quantized plateaus of $\theta$ term, which are mainly determined by the Landau level structures on the top and bottom surfaces.
\end{abstract}

\pacs{73.43.$-$f, 85.75.$-$d, 85.70.Kh}

\maketitle

\section{Introduction} 

After a recent discovery of  topological insulators (TIs) \cite{HasanRMP2010,QiRMP2011}, the quest for numerous fascinating effects in response to external perturbations in TIs has started. One of them is the magnetoelectric (ME) effect, which is a long standing issue in multiferroics \cite{Tokura2007} and has been recently introduced in TIs \cite{HasanRMP2010,QiRMP2011,Qi2008,Essin2009,Vazifeh2010,Nomura2011,Tse2010}. The essence of the ME effect is that the external electric or magnetic fields induce magnetization or polarization in a TI, respectively. From the effective topological field theory predictions \cite{Qi2008}, the ME effect in three-dimensional (3D) TIs can be described by introducing a new term $U_{\theta}$ in the energy density for the axion electrodynamics \cite{HasanRMP2010,QiRMP2011,Qi2008,Essin2009,Chen2011,Nomura2011,Tse2010,Wilczek1987, Vazifeh2010,Coh2011,Sitte2012}:
\begin{equation}
\label{Eq:Enden}
U_{\theta}=-\frac{e^2}{4\pi^2\hbar c}\theta\mathbf{E}\cdot\mathbf{B},
\end{equation}
where $\mathbf{E}$ and $\mathbf{B}$ are the electric and magnetic fields respectively. The $\theta$ term, $U_{\theta}$ \cite{Wilczek1987}, characterizes the 3D TIs. In Eq.~(\ref{Eq:Enden}), parameter $\theta$ takes values $\pm\pi$  for TIs, whereas it is $0$ for ordinary insulators.
The magnetization $\mathbf{M}$ and electric polarization $\mathbf{P}$ are obtained from the energy density as follows:
\begin{eqnarray}
\mathbf{M}&=&-\frac{\partial U_{\theta}}{\partial \mathbf{B}}=\frac{e^2 }{4\pi^2\hbar c}\theta \mathbf{E},\\
\qquad \mathbf{P}&=&-\frac{\partial U_{\theta}}{\partial\mathbf{E}}=\frac{e^2}{4\pi^2\hbar c}\theta\mathbf{B}, 
\end{eqnarray}
and clearly show the cross-correlated responses \cite{Tokura2007,Qi2008,Essin2009,Essin2010,Malashevich2010,Malashevich2011}.

In a strong magnetic field, which breaks time reversal symmetry, however, a controversy still remains in the literature regarding the quantization of ME responses. According to Ref.~\onlinecite{Tse2010}, which considers only TI surface states, the response should be quantized: $\theta/\pi=2N+1$ ($N$ being an integer) when the electron densities on the top and bottom surfaces are balanced, i.e., no gate voltage is applied. On the other hand, Ref.~\onlinecite{Coh2011}, considering only bulk states in a magnetic field, concludes that $\theta$ is not quantized but rather arbitrary. 

\begin{figure}
\includegraphics[width=0.99\columnwidth]{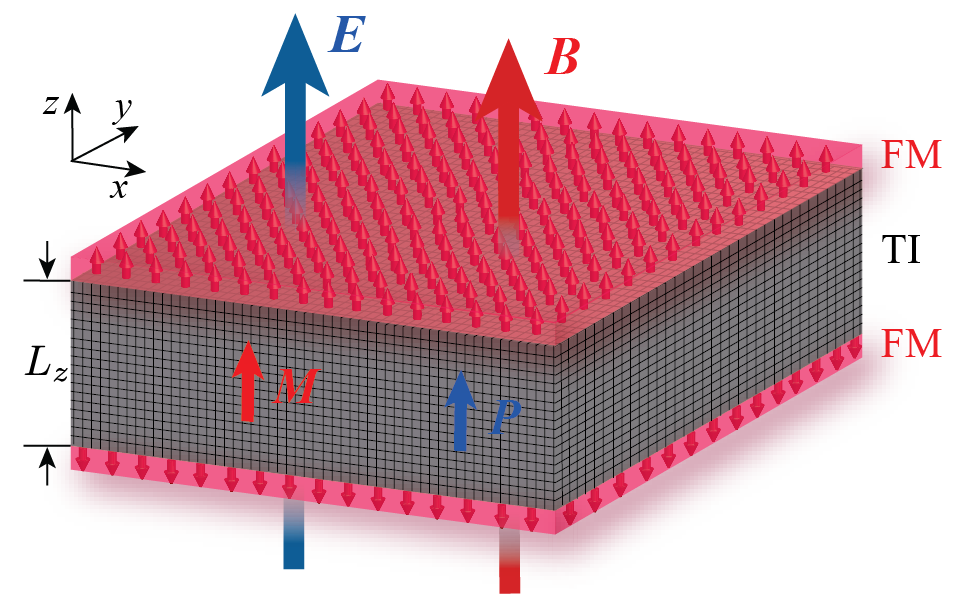}
 \caption{(Color online) The ME effect in a 3D TI of slab geometry sandwiched between two ferromagnets (FMs) with opposite magnetizations.}
 \label{fig:EM}
\end{figure}

In this paper, to resolve this controversy, we study the effect of an ultrahigh magnetic field on TIs. To fully treat both bulk and surfaces, we consider the lattice Hamiltonian \cite{Zhang2009,Liu2010} for a 3D TI in a slab geometry, sandwiched between two thin ferromagnets with magnetizations pointing in the opposite directions along $z$ axis, see Fig.~\ref{fig:EM}. The role of ferromagnetic films can also be played by an appropriate doping of the TI with magnetic impurities. 
We first reproduce the $\theta$ term (\ref{Eq:Enden}) with $\theta=\pm\pi$, in the balanced case, from the total energy of electrons. This means that the axion electrodynamics remains unchanged as long as the bulk gap is not destroyed, even though time-reversal symmetry, which plays essential role in TIs, is broken by the magnetic field. However, as the TI thickness is decreased, we observe that the ME effect starts to vanish. Based on this result, we estimate the critical thickness of TI film capable of displaying the robust ME effect, which can be useful for future experimental studies, especially in the view of recent experimental progress on TI thin films \cite{Checkelsky2012,Chang2013}. Moreover, we also study the case of both strong electric and magnetic fields. We show that the response to a small change in the electric field can be characterized in this regime by different $\theta$, which are \textit{still quantized} as long as the Fermi level resides within the bulk gap. We show that the quantization rule of $\theta$ is determined by Landau level structures on the top and bottom surfaces, while the bulk states do not play an important role in thick TI films.

\section{Model}

To study the ME effect in TIs, we employ the Wilson-Dirac tight-binding Hamiltonian 
for 3D lattice TI model \cite{Zhang2009,Liu2010}, which is a simplified version of the effective four-band Hamiltonian for newly discovered 3D TIs such as $\mathrm{Bi}_2\mathrm{Se}_3,\mathrm{Bi}_2\mathrm{Te}_3$, and $\mathrm{Sb}_2\mathrm{Te}_3$ \cite{Xia2009,Chen2009,Hsieh2009}. We introduce external magnetic field $\mathbf{B}$, applied in $z$ direction, through the Peierls substitution and set its magnitude $B$ so that the magnetic flux going through a single plaquette is a rational fraction of the magnetic flux quantum. In other words, $B=(p/q)\Phi_0/a^2$, where $p$ and $q$ are natural numbers with $p<q$, $\Phi_0=hc/e$ is the magnetic flux quantum, and $a$ is the lattice constant. For convenience, we set $e=c=\hbar=a=1$ in the following so the magnetic field becomes $B=2\pi p/q$. Using Landau gauge $\mathbf{A}=(0,Bx,0)$, the Hamiltonian is defined as 
\begin{equation}		
\label{eq:Wlsn}	
H_0=-\frac{1}{2}\sum_{\mu=x,y,z}(T_{\mu}^{}+T_{\mu}^{\dag})+(m+3r)\sum_{\mathbf{R}} c^\dag_{\mathbf{R}}\beta c_{\mathbf{R}}. 
\end{equation}
Here, the translation operators in $\mu=x,y,z$ directions are
\begin{equation}
\label{eq:Wlsn2}
T_{\mu}=\sum_{\mathbf{R}}c^\dag_{\mathbf{R}+\mathbf{a}_{\mu}}e^{i A_{\mu}}(r\beta-it\alpha_{\mu})c_{\mathbf{R}},
\end{equation}
where $t$ and $r$ are the hopping parameters with and without spin flip, respectively, $m$ is the mass characterizing the spin-orbit interaction, $\mathbf{R}=(n_x,n_y,n_z)$ is the lattice coordinate, $\mathbf{a}_{\mu}$ is the unit lattice vector in $\mu$ direction, and $\alpha_{\mu}$, $\beta$ are the standard Dirac matrices:
\begin{equation*}
  \alpha_\mu=
  \begin{pmatrix}
    0&\mathbf{\sigma_\mu}\\
    \mathbf{\sigma_\mu}&0
  \end{pmatrix},\quad\beta=
  \begin{pmatrix}
    I&0\\
    0&-I
  \end{pmatrix},
\end{equation*}
where $\sigma_{\mu}$ are the Pauli matrices. 

To take into account the effect of an external electric field and the interaction of electron spins on TI surfaces with magnetization, we introduce additional terms $H_E$ and $H_s$, respectively. Then, the total Hamiltonian is
\begin{eqnarray}
 \label{eq:Ham}
  H&=&H_0+H_E+H_s,\\
  H_E&=&\sum_{\mathbf{R}}c^\dag_{\mathbf{R}}U(n_z)
  \begin{pmatrix}
    I&0\\
    0&I
  \end{pmatrix}
  c_{\mathbf{R}}
\end{eqnarray}
with the potential energy $U(n_z)=-E(n_z-L_z/2)$, where $L_z$ is the TI film thickness. The surface spin contribution \cite{Qi2008, Nomura2011, LiuPRL2009, Abanin2011}, see Fig.~\ref{fig:EM}, is given by
\begin{equation}
H_s=\sum_{\mathbf{R}}c^\dag_{\mathbf{R}} b(n_z) \Sigma_s^+c_{\mathbf{R}}, \,
b(n_z)=
\begin{cases}
b_s,&n_z=L_z,\\
-b_s,&n_z=0.
\end{cases}
\end{equation}
Here, $\Sigma_s^+=\mathrm{diag}(\mathbf{\sigma}_z,\mathbf{\sigma}_z)$ and $b_s$ is the surface spin constant. 
    
\section{Results and Discussion} 
We investigate the magnetoelectric effect by calculating the energy eigenvalues of electrons. Since our model has translational invariance in $x$ and $y$ directions, Eq.~(\ref{eq:Ham}) can be written as 
\begin{equation}
 H=\sum_{k_x,k_y,n_z}c^{\dag}_{n_z}(k_x,k_y){\cal H}_{n_z,n_z'}(k_x,k_y)c_{n_z'}(k_x,k_y),
\end{equation}
where ${\cal H}_{n_z,n_z'}(k_x,k_y)$ is $4q(L_z+1)\times 4q(L_z+1)$ matrix, see Appendix for details. By numerically performing an exact diagonalization of this Hamiltonian matrix, we solve the eigenvalue equation 
${\cal H}(k_x,k_y)|\lambda,k_x,k_y\rangle=\epsilon_{\lambda}(k_x,k_y)|\lambda,k_x,k_y\rangle$.
The ME effect is described by the energy density of electrons:
\begin{equation}
 U_{\rm tot}=\frac{1}{\mathcal{V}}\sum_{k_x,k_y}{\sum_{\lambda}}'\epsilon_{\lambda}(k_x,k_y),
\end{equation}
where $\mathcal{V}$ is the TI volume and $\sum_{\lambda}'$ is the sum of eigenvalues below the Fermi level, which is fixed at $\epsilon_F=0$ in the following.
In the presence of both electric $\mathbf{E}$ and magnetic $\mathbf{B}$ fields, $U_{\rm tot}$ has a term $\propto\mathbf{E}\cdot\mathbf{B}$ as expected from the axion electrodynamics, Eq.~(\ref{Eq:Enden}).

\begin{figure}
\includegraphics[width=0.97\columnwidth]{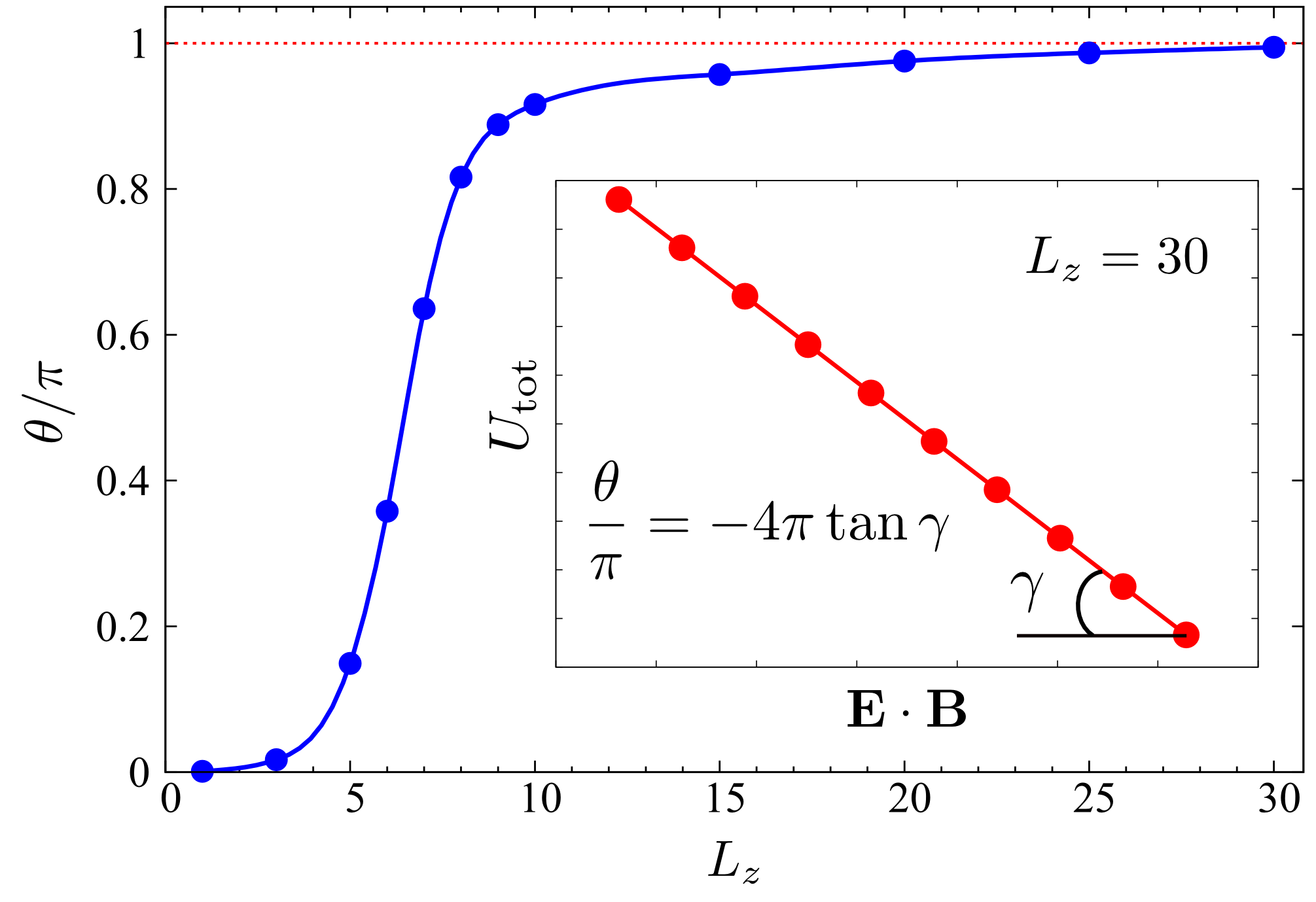}
\caption{(Color online) Parameter $\theta$ as a function of TI film thickness $L_z$. The inset shows the energy density of electrons in a weak electric field and its linear relationship to the product of electric and magnetic fields for $L_z=30$.}
\label{fig:thetaLz}
\end{figure}

The energy density for a typical case is plotted in the inset of Fig.~\ref{fig:thetaLz}. For fixed magnetic field, we change the electric field and plot the energy density as a function of $\mathbf{E}\cdot\mathbf{B}$.
From Eq.~(\ref{Eq:Enden}), we estimate the parameter $\theta$ by $\theta=-4\pi^2\tan\gamma$, where $\tan\gamma$ is the slope of the linear dependence in the inset of Fig.~\ref{fig:thetaLz}. In all our calculations, we set the hopping parameters $t=r=1$, the mass $m=-1$, and choose the number of layers in $x$ direction $L_x$, so that the number of points in $k_x$ direction of the magnetic Brillouin zone is $L_x/q=32$.

\begin{figure}
\centering 
\includegraphics[width=1\columnwidth]{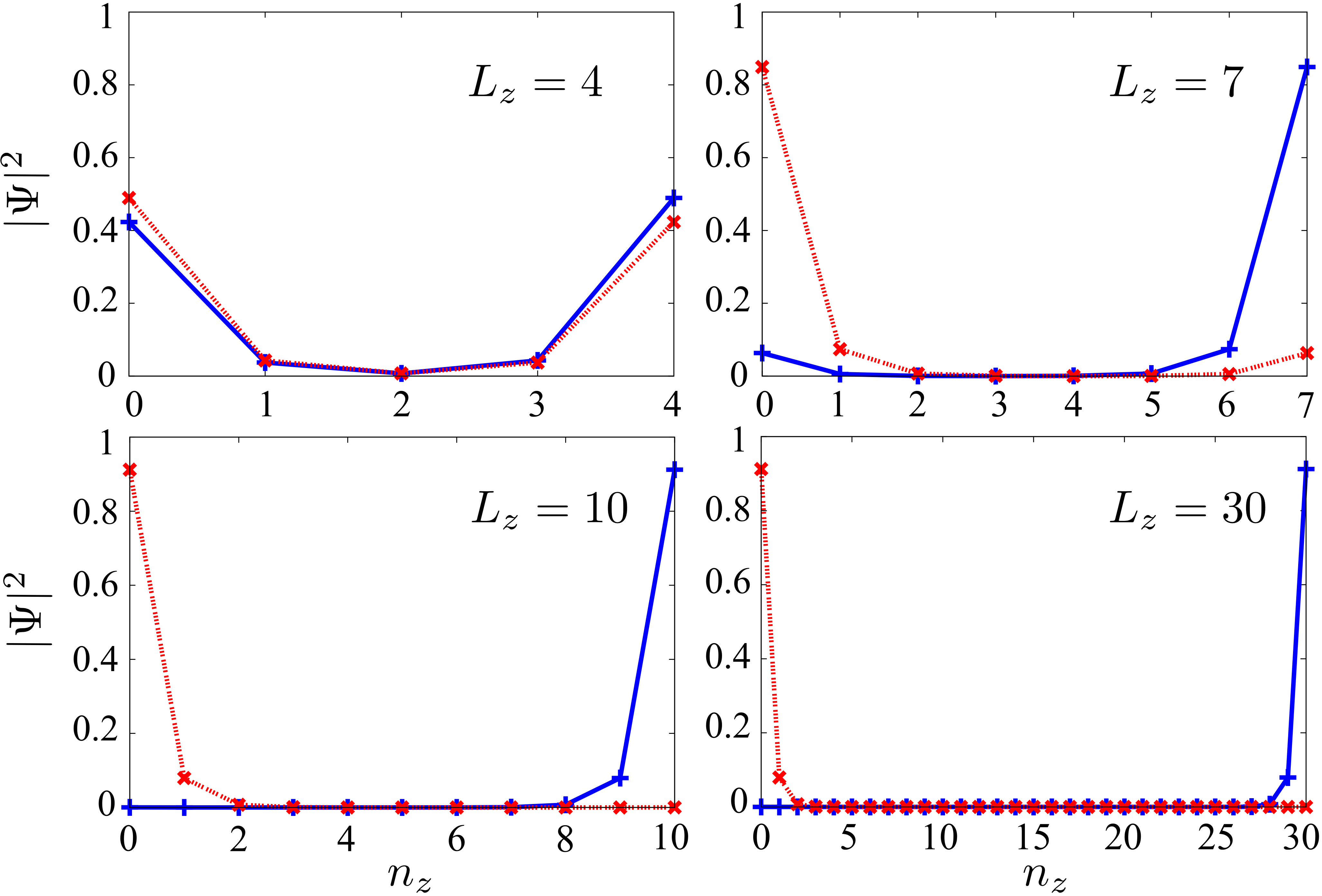}
\caption{(Color online) The hybridization of surface wave functions $\Psi$ for TI thicknesses $L_z=4,7,10$, and $30$. The red dashed line corresponds to the bottom TI surface wave function and the blue solid line corresponds to the top surface wave function.}
\label{fig:LzComb}
\end{figure}

First, to verify our method, we study $\theta$ as a function of TI thickness $L_z$ for the fixed magnetic field $B=\pi/5$ corresponding to $p=1$ and $q=10$, see the main panel of Fig.~\ref{fig:thetaLz}. The data show that $\theta$ is close to $\pi$ at large $L_z$, whereas it decreases notably below $L_z\approx10$. As $L_z$ approaches 1, the parameter $\theta$ almost vanishes, which means that in very thin TI films the ME effect disappears. When the TI thickness is very small, it is expected that its surface wave functions hybridize with each other \cite{Linder09,Essin2009}. To confirm it, we calculate the spacial profiles of the wave functions in $z$ direction for the surface states.  At the $\Gamma$ point $(k_x=k_y=0)$, they are calculated for several values of $L_z$ and shown in Fig.~\ref{fig:LzComb}. We observe that for $L_z=30$ the wave functions are strongly localized near the TI surfaces, whereas overlap of these wave functions increases with decreasing TI thickness. At $L_z=4$ the top and bottom wave functions almost completely overlap. This clearly indicates that the reduction of the ME effect is associated with the hybridization of the surface wave functions. From Fig.~\ref{fig:thetaLz}, the value of $L_z$ below which the ME effect starts to significantly diminish can be estimated as $\sim 10$. For typical TIs \cite{Cui2004,Hsieh2008,Chen2009,Xia2009,Hsieh2009}, this value corresponds to $\sim 10$ nm. 

\begin{figure}
\centering
\includegraphics[width=1\columnwidth]{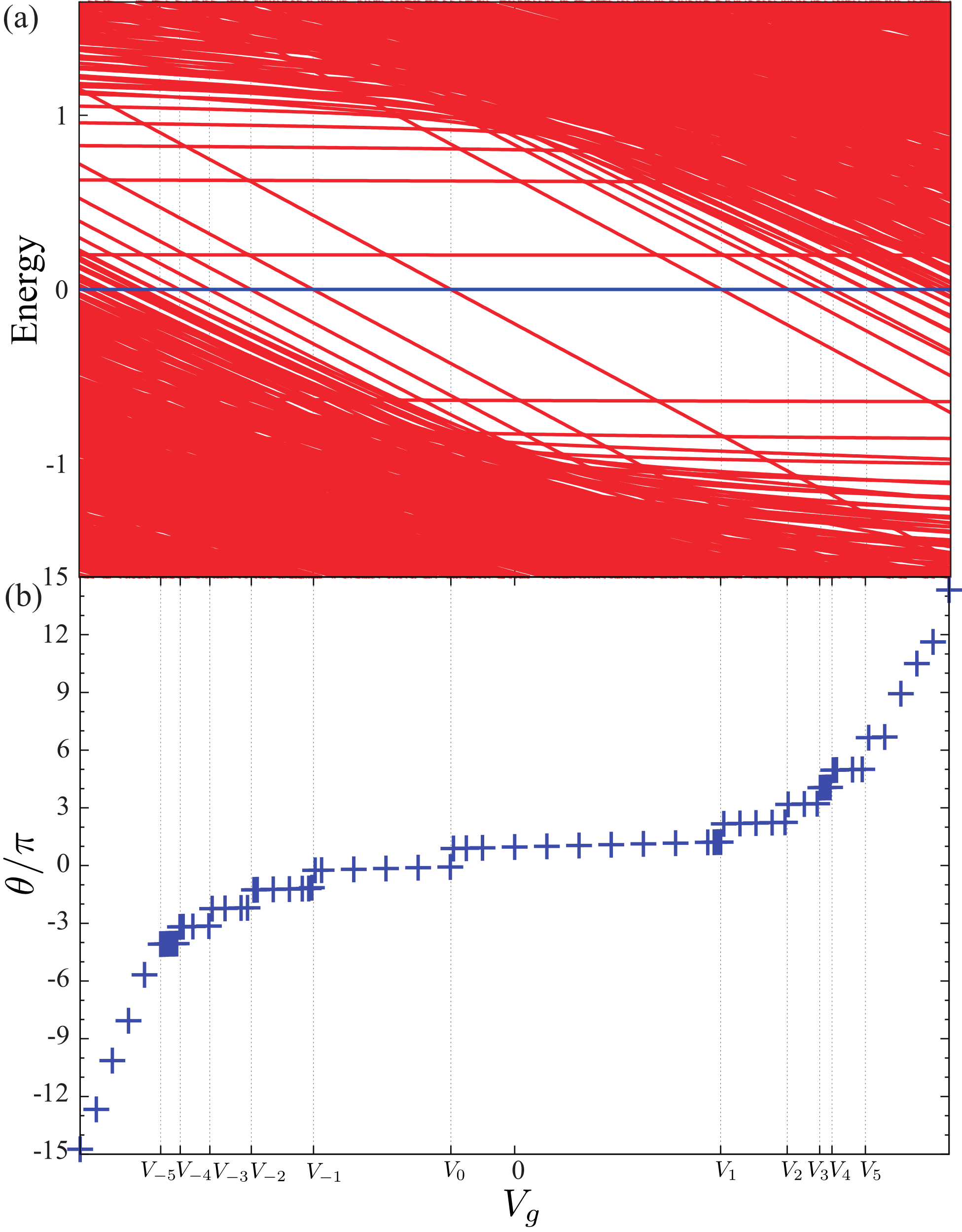}
\caption{(Color online) (a) The TI energy levels (at $\Gamma$ point) as a function of gate voltage $V_g$ measured in units of $E_{g}L_z$. The surface Landau levels cross the Fermi level. The red lines represent the surface Landau levels and the blue line is the Fermi level. (b) The ME response quantization in a TI vs $V_g$. }
\label{fig:EgLz}
\end{figure}

So far, we have considered a TI in a weak electric field. Experimentally, however, ultrahigh electric fields can be achieved by ionic liquid gating \cite{Checkelsky2012}, and it is therefore interesting to study a nontrivial response of a TI to both strong electric and magnetic fields.  In the presence of a strong magnetic field, Landau levels form on top and bottom TI surfaces. The existence of well-defined Landau levels in TIs has been confirmed by recent experimental observations of the quantum Hall effect \cite{Brune2011,Chang2013}. In our simulation, we fix for simplicity the Landau levels on the bottom surface, whereas the gating induced electric field, $E_g$, shifts the energy levels on the top surface. As the gate voltage changes from $V_g\equiv E_gL_z=0$, the Landau level energies on the top surface move. Figure~\ref{fig:EgLz}(a) shows that Landau levels cross the Fermi level
at the following gate voltages: $V_{-5},\,V_{-4},\,V_{-3},\ldots,V_{5}$. In the nonlinear regime of strong electric or magnetic fields, the axion electrodynamics corresponding to $\theta=\pm\pi$ in general is not expected to be applicable. To probe this regime, we vary the electric field by small $\delta \mathbf{E}$ near $\mathbf{E}_g$, i. e. the total electric field is $\mathbf{E}=\mathbf{E}_g+\delta\mathbf{E}$, and study the response to $\delta\mathbf{E}\cdot\mathbf{B}$. 
Then $\delta\mathbf{E}$ has a role of $\mathbf{E}$ in Eq.~(\ref{Eq:Enden}) and $\theta(E_g,B)$ is defined as the overall coefficient describing the ME effect response.

Figure~\ref{fig:EgLz}(b) shows $\theta/\pi$ as a function of $V_g$. One can see that there are several plateaus where $\theta/\pi$ is approximately quantized in integers $\{-4,-3,-2,-1,0,1,2,3,4,5\}$. Every new quantization plateau in Fig.~\ref{fig:EgLz}(b) occurs at the points where Landau levels cross the Fermi level.
The exact quantization would be in agreement with the prediction of Ref.~\onlinecite{Tse2010} for weak electric fields. 
As we show below, the deviation from the exact quantization is mostly attributed to the asymmetry of TI surface wave functions due to the application of the large gate voltage. Thus, Fig.~\ref{fig:EgLz}(b) shows an important fact that although time-reversal symmetry is broken in a 3D TI by a strong magnetic field, $\theta$ is not arbitrary as it has been claimed in Refs.~\onlinecite{Coh2011,Sitte2012}, but rather quantized as long as the Fermi level is still within the bulk gap. In accord with this, the two continuous regions of the dependence in Fig.~\ref{fig:EgLz}(b) at the left and right edges correspond to the cases when the Fermi level enters continuous (valence or conduction) bands.  In the two continuous regions, however, our results only qualitatively show the behavior of the parameter $\theta$, because we do not consider any screening effects coming from conduction and valence bands that may suppress the $\theta$-term, and also because beyond the range bounded by $V_{-5}$ and $V_{5}$, see Fig.~\ref{fig:EgLz}(a), there are multiple crossings of Landau levels very close to each other. 

\begin{figure}
\centering 
\includegraphics[width=0.97\columnwidth]{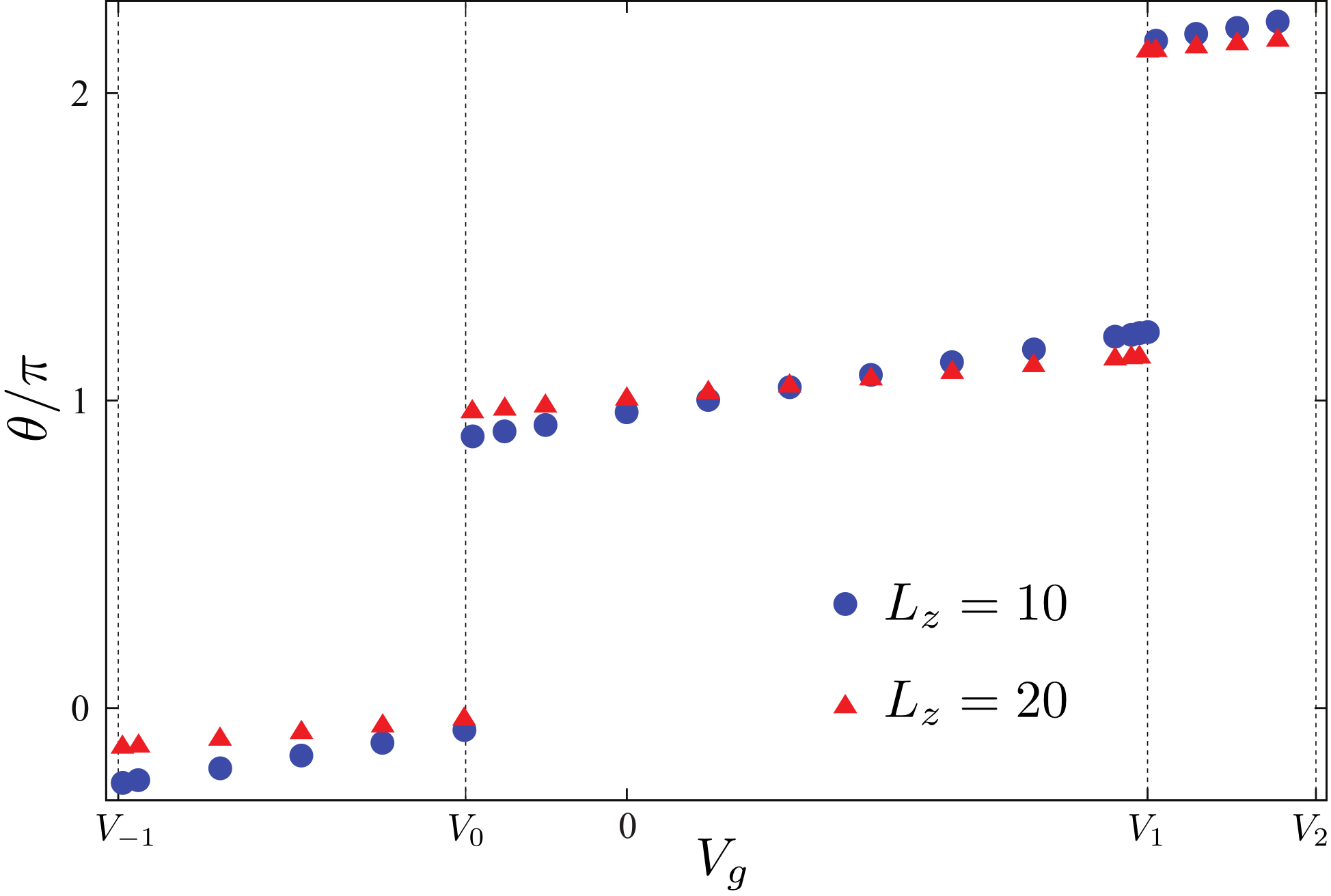}
\caption{(Color online) The quantization of $\theta$ as a function of gate voltage $V_g$ for $L_z=10$ (blue circles) and $L_z=20$ (red triangles). The comparison shows that when the TI thickness $L_z$ increases two times the slope of each quantization plateau becomes two times smaller.}
\label{fig:Lz10vsLz20}
\end{figure}

To investigate the hypothesis that the lack of exact quantization in $\theta$ is attributed to the asymmetry of TI surface wave functions due to gating, we repeat the same calculation for a thicker TI film. The details of the quantization of $\theta$ as a function of gate voltage $V_g$ are shown in Fig.~\ref{fig:Lz10vsLz20} for two different TI film thicknesses. The results for $L_z=10$ are shown by blue circles and for $L_z=20$ by red triangles. At a fixed gate voltage $V_g$ ($=E_{g}L_{z}$), applied electric field for $L_z=10$ case is two times stronger than that for $L_z=20$. The comparison shows that when $L_z$ increases two times the slope of each quantization plateau becomes two times smaller. This indicates that for thick enough TI films the plateaus should become horizontal, because in this case the asymmetry of surface wave functions can be disregarded since their overlap is too small. As a result, for the thick films, $\theta$ depends only on the surface states, but not the bulk states. When the bulk gap is not destroyed, $\theta$ is determined by the surfaces and thus quantized as the Hall conductivity is quantized, i.e.,
\begin{equation}
\frac{\theta}{\pi} = \nu_t -\nu_b,
\end{equation}
where $\nu_{t(b)}$ are the Landau level filling factors for the top (bottom) TI surfaces. In the case of zero applied gate voltage,  $\nu_{t}=-1/2$, $\nu_{b}=1/2$ and therefore $|\theta/\pi|=1$, c.f. Fig.~\ref{fig:thetaLz}. 

By applying a magnetic field to the TI model, Eq.~(\ref{eq:Wlsn}), we break time-reversal symmetry and as a result the system crosses from symmetry class AII, where $\theta$ is quantized $0$ or $\pi$, to class A where in general $\theta$ can be arbitrary. Thus we show that depending on the perturbations (orbital or Zeeman magnetic fields or electric field), $\theta$ is still quantized in class A for our model. The quantization rules can be rather nontrivial depending on a combination of perturbations, however, we can successfully explain these rules based on the Landau level physics.

Finally, we study the effects of bulk magnetic interaction on the ME effect. The Zeeman type interaction between the electron spin and applied magnetic field is described by 
\begin{eqnarray}
\label{minus}
\mathcal{H}'&=&\mu_BB_z
\begin{pmatrix}
g_1\sigma_z&0\\
0&g_2\sigma_z
\end{pmatrix},
\end{eqnarray}
where $g_1$ and $g_2$ are the effective $g$ factors depending on the TI materials. In most of cases, for materials such as Bi$_2$Se$_3$ and Bi$_2$Te$_3$, $g_1$ and $g_2$ have opposite signs \cite{Liu2010}. Therefore, as a simplified model, we consider the case of $g_2=-g_1$. In addition, we study the influence of the exchange interaction between the electron spins and magnetic impurities that can be introduced by the appropriate doping of the TI bulk. Within the virtual crystal approximation, this interaction has the form
\begin{eqnarray}
\label{plus}
\mathcal{H}'&=&J\overline{M}_z
\begin{pmatrix}
\sigma_z&0\\
0&\sigma_z
\end{pmatrix},
\end{eqnarray}
where $J$ is the exchange interaction constant and $\overline{M}_z$ is the mean value of magnetic moments in the bulk. These two interactions, Eqs.~(\ref{minus}) and (\ref{plus}), can be covered by two simple model Hamiltonians: $\mathcal{H}'=b_b\Sigma_z^{\pm}$ with $\Sigma_z^{\pm}=\mathrm{diag}(\mathbf{\sigma}_z, \pm\mathbf{\sigma}_z)$ and $b_b$ being the bulk spin constant. The sign ``$+$'' in $\Sigma_z^+$ indicates bulk electrons coupling with magnetization, while ``$-$'' sign indicates coupling with the external magnetic field.  

\begin{figure}
\centering 
\includegraphics[width=0.97\columnwidth]{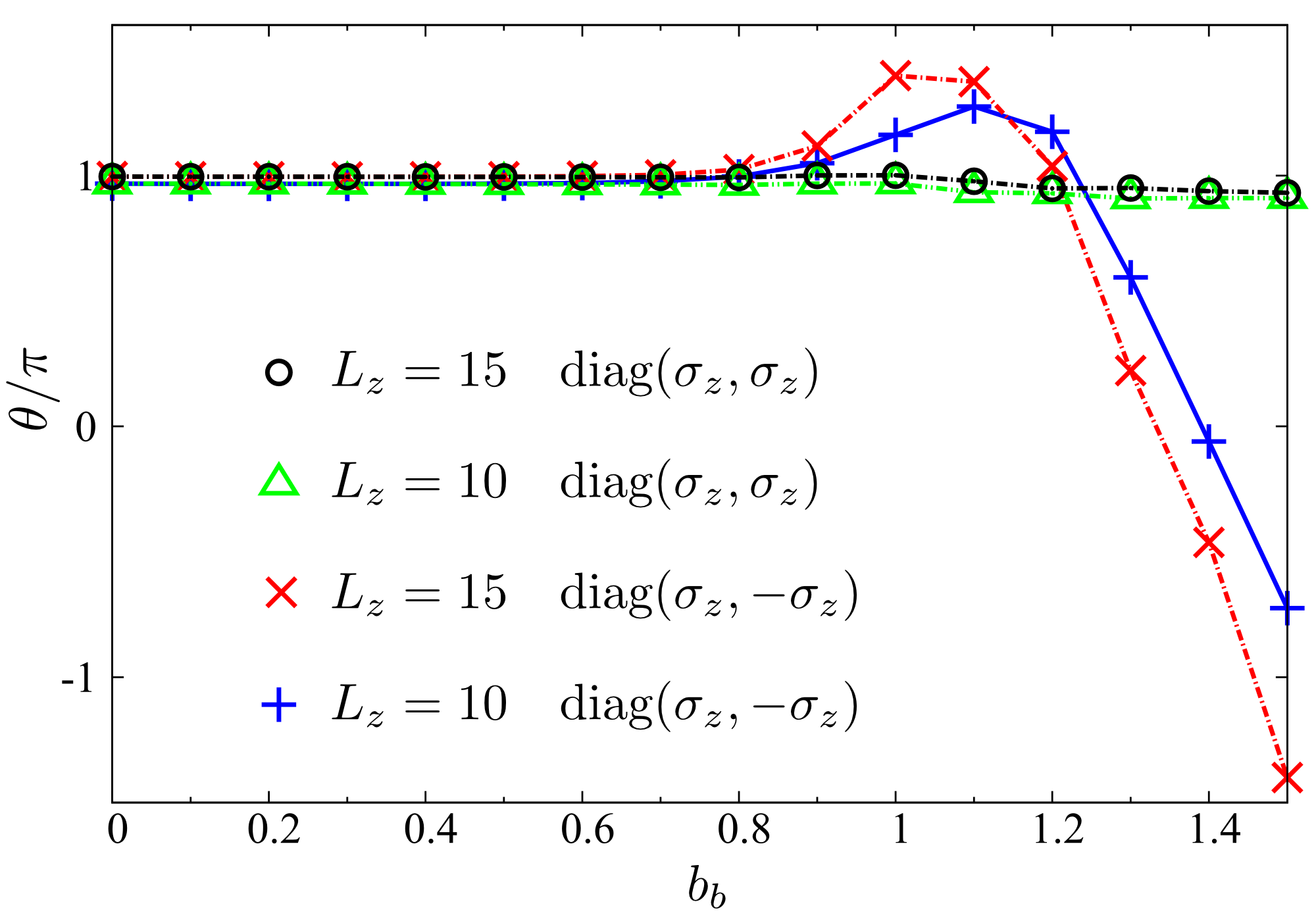}
\caption{(Color online) The bulk spin interaction influence on the ME effect. The triangles and circles represent $\Sigma_z^+=\mathrm{diag}(\mathbf{\sigma}_z,\mathbf{\sigma}_z)$ case, whereas ``$\times$'' and ``$+$'' symbols represent $\Sigma_z^-=\mathrm{diag}(\mathbf{\sigma}_z,-\mathbf{\sigma}_z)$ case.}
\label{fig:bulkspin}
\end{figure}

The calculation results for magnetic field $B=\pi/10$ are shown in Fig.~\ref{fig:bulkspin}, where the red and blue curves correspond to $\Sigma_z^-=\mathrm{diag}(\mathbf{\sigma}_z,-\mathbf{\sigma}_z)$, red is for $L_z=15$ and blue is for $L_z=10$.  The black and green curves correspond to $\mathrm{diag}(\mathbf{\sigma}_z,\mathbf{\sigma}_z)$ case, with the black for $L_z=15$ and green for $L_z=10$. One can see from Fig.~\ref{fig:bulkspin} that $\theta/\pi=1$ persists approximately up to $b_b=1$, which corresponds to the TI regime. At $b_b=|m|=1$, the system crosses to the Weyl semimetal regime \cite{Wan2011,Burkov2011} for $\mathrm{diag}(\mathbf{\sigma}_z,\mathbf{\sigma}_z)$ case and therefore no quantization of $\theta$ beyond $b_b=1$ is expected \cite{Zyuzin2012}. These results show remarkable robustness of the ME effect in TI phase even for very strong magnetic fields or high concentration of magnetic impurities that affect the bulk structure of the TI. It seems to be irrespective of microscopic details of coupling with the magnetic impurities or field, since it works equally well in the TI regime for the toy models with both $\pm$ signs.
 
\section{Summary}

We have shown that the ME effect in TIs survives even in both strong magnetic and electric fields even though time-reversal symmetry is broken. We have described the ME effect  beyond the linear response regime and have confirmed that it is robust but takes a fascinating form of multiple quantization related to Landau levels crossing of the Fermi level. In addition, the influence of the TI surface hybridization has been studied in very thin TI films, describing how the ME effect vanishes in the limit of very thin TIs.  Furthermore, the ME effect is shown to persist even for strong bulk interactions with magnetic field or magnetic impurities.  

\acknowledgments 

We would like to thank the staff of the Center for Computational Materials Science at the IMR, Tohoku University for their support of the supercomputing facilities. K.N. was supported by Grant-in-Aid for Scientific Research (Nos. 24740211 and 25103703) from the Ministry of Education, Culture, Sports, Science and Technology (MEXT), Japan. O.A.T. acknowledges support  by the Grants-in-Aid for Scientific Research (Nos. 25800184 and 25247056) from the MEXT, Japan.

\appendix

\setcounter{secnumdepth}{0}
\section{APPENDIX: MODEL HAMILTONIAN IN MOMENTUM SPACE}
\label{append}

We present below the details of diagonalization of Wilson-Dirac Hamiltonian \cite{Zhang2009,Liu2010} given by Eqs.~(\ref{eq:Wlsn}) and (\ref{eq:Wlsn2}) in the main text. We use a slab geometry with periodic boundary conditions in $x$ and $y$ directions and fixed boundary condition in $z$ direction, $0\le n_z\le L_z$ where $L_z +1$ is the number of lattice sites in $z$ direction. According to our choice of the magnetic field and Landau gauge, there is $q$ periodicity in the Aharonov-Bohm (AB) phase, i.e., the lattice sites whose $x$ coordinate is $\pmod{q}$ have the same AB phase. Therefore we can define a new unit cell that is $q$ times larger than the original one by introducing an additional coordinate $s$, 	   
\begin{equation}
  n_x = qn'_x +s,\quad 1\le s\le q, 
\end{equation}
where $n'_x$ is a coordinate of a one-dimensional unit cell in $x$ direction, containing $q$ lattice points and $s$ is the coordinate of the lattice site within the unit cell. As a result, the lattice can be represented by $\mathbf{R}=(s,n'_x,n_y,n_z)$, where $(n'_x,n_y,n_z)$ corresponds to a coordinate of a 3D unit cell and $s$ corresponds to a coordinate of a lattice point within that unit cell.

Because of periodicity in $x$ and $y$ directions, it is convenient to introduce a new annihilation (and corresponding creation) operator $c_{s,n_z}(\mathbf{k})$ by means of Fourier transformation:
\begin{equation}
  \label{eq:Ann}
  c_{s,n'_x,n_y,n_z} = \sqrt{\frac{q}{L_x L_y}} \sum_{k_x,k_y} e^{ik_xqn'_x} 
  e^{ik_yn_y} c_{s,n_z}(\mathbf{k}),
\end{equation}
where $L_x$ and $L_y$ are the numbers of layers in $x$ and $y$ directions, respectively, and $\mathbf{k}=(k_x,k_y)$ is the wave vector in $xy$ plane. In Eq.~(9) of the main text we omitted for brevity the subscript $s$ in $c_{s,n_z}(\mathbf{k})$.   In terms of these operators the Hamiltonian (\ref{eq:Wlsn}) becomes
\begin{equation}
   \label{eq:waven}
  H_0 = \sum_{\mathbf{k}} c^\dag(\mathbf{k})\mathcal{H}_0(\mathbf{k})c(\mathbf{k}),
\end{equation}
where 
  $c^\dag (\mathbf{k})=(c^\dag_{10},c^\dag_{11},\ldots,c^\dag_{1L_z},\ldots,c^\dag_{q0},c^\dag_{q1},\ldots,c^\dag_{qL_z})$ and the matrix $\mathcal{H}_0(\mathbf{k})$ is
\begin{equation}
 \mathcal{H}_0(\mathbf{k}) =
  \begin{pmatrix}
    \Delta_1 & \Xi^\dag & \phantom{0} & \phantom{0} & \Omega \\
    \Xi & \Delta_2 & \Xi^\dag & \phantom{0} & \phantom{0} \\
    \phantom{0} & \Xi & . & . \\
    \phantom{0} & \phantom{0} & . & . & \Xi^\dag \\
    \Omega^\dag & \phantom{0} & \phantom{0} & \Xi & \Delta_q 
  \end{pmatrix}.
\end{equation}
Here the diagonal block elements are
\begin{equation*}
  \Delta_{\lambda} =
  \begin{pmatrix}
    a_{\lambda} & b^\dag & \phantom{0} & \phantom{0} & \phantom{0} \\
    b & a_{\lambda} & b^\dag \\
    \phantom{0} & b & . & . \\
    \phantom{0} & \phantom{0} & . & . & b^\dag \\
    \phantom{0} & \phantom{0} & \phantom{0} & b & a_{\lambda} 
  \end{pmatrix}
\end{equation*}
with $b = (it\alpha_z -r\beta)/2$ and $a_{\lambda} = t\sin(k_y-2\pi p\lambda /q)\alpha_y+[m+3r-r\cos(k_y-2\pi p\lambda/q)]\beta.$ The nonzero off-diagonal block elements of $\mathcal{H}_0 (\mathbf{k})$ are
$\Xi = (it\alpha_x -r\beta)I_{L_z+1}/2$ and $\Omega = (it\alpha_x -r\beta)e^{-ik_x q}I_{L_z+1}/2$, where $I_{L_z+1}$ is  $(L_z+1)\times (L_z+1)$ unit matrix.

To include the external electric field $H_E$ as well as surface spin interaction with magnetization $H_s$ terms, we apply the same transformation given by  Eq.~(\ref{eq:Ann}). Then the total model Hamiltonian, $H = H_0 + H_E + H_s$, is represented as $\sum_{\mathbf{k}} c^\dag(\mathbf{k})\mathcal{H}(\mathbf{k})c(\mathbf{k})$ with
\begin{equation} 
  \mathcal{H}(\mathbf{k}) =
  \begin{pmatrix}
    \Delta_1+\Lambda & \Xi^\dag & \phantom{0} & \phantom{0} & \Omega \\
    \Xi & \Delta_2+\Lambda & \Xi^\dag & \phantom{0} & \phantom{0} \\
    \phantom{0} & \Xi & . & . \\
    \phantom{0} & \phantom{0} & . & . & \Xi^\dag \\
    \Omega^\dag & \phantom{0} & \phantom{0} & \Xi & \Delta_q+\Lambda
  \end{pmatrix}.
\end{equation}
Here the electric field and overall spin contribution coming from both TI surfaces are given by matrix $\Lambda = \mathrm{diag} (g_0, g_1,\dots, g_{L_z})$ with $g_{n_z} = U(n_z) I_4 +b(n_z)\Sigma_s^{+}$ where $I_4$ is $4\times4$ unit matrix.

\bibliography{TI}

\end{document}